\RequirePackage{fix-cm}

\documentclass[twocolumn]{svjour3}          

\usepackage[T1]{fontenc}
\usepackage[sort,compress,numbers]{natbib}
\usepackage{graphicx}
\usepackage{bm}
\usepackage{epstopdf}
\usepackage[colorlinks=true,linkcolor=blue,citecolor=blue]{hyperref}

\journalname{J Supercond Nov Magn}

\newcommand{\arttitle}[1]{#1.}
\newcommand{\linkDOIOK}[1]{\unskip. doi:\href{http://dx.doi.org/#1}{#1}}

\begin{document}

\title{Phase transitions in {\it quasi}-one dimensional system with unconventional superconductivity
}

\titlerunning{Phase transitions in {\it quasi}-one dimensional system \ldots}

\author{Andrzej Ptok \and Agnieszka Cichy \and Karen Rodr\'{i}guez \and Konrad Jerzy Kapcia}


\authorrunning{A. Ptok, A. Cichy, K. Rodr\'{i}guez, K.J. Kapcia} 

\institute{A. Ptok \at
              Institute of Nuclear Physics, Polish Academy of Sciences, ul. E. Radzikowskiego 152, PL-31342 Krak\'{o}w, Poland \\
              Institute of Physics, Maria Curie-Sk\l{}odowska University, Plac M. Sk\l{}odowskiej-Curie 1, PL-20031 Lublin, Poland \\
              \email{aptok@mmj.pl} \\
           \and
           A. Cichy \at
	          Solid State Theory Division, Faculty of Physics, Adam Mickiewicz University in Pozna\'{n}, ul. Umultowska 85, PL-61614 Pozna\'{n}, Poland \\
              Institut f\"{u}r Physik, Johannes Gutenberg-Universit\"{a}t Mainz, Staudingerweg 9, D-55099 Mainz, Germany \\
              \email{agnieszkakujawa2311@gmail.com} \\
            \and
            K. Rodr\'{i}guez \at
              Departamento de F\'{i}sica, Universidad del Valle, A.A. 25360, Cali, Colombia \\
              Centre for Bioinformatics and Photonics -- CiBioFi, Calle 13 No. 100-00, Edificio 320 No. 1069, Cali, Colombia \\
              \email{karem.c.rodriguez@correounivalle.edu.co} \\
              \and
            K. J. Kapcia \at
              Institute of Nuclear Physics, Polish Academy of Sciences, ul. E. Radzikowskiego 152, PL-31342 Krak\'{o}w, Poland \\
              Institute of Physics, Polish Academy of Sciences, Aleja Lotnik\'{o}w 32/46, PL-02668 Warsaw, Poland \\
              \email{konrad.kapcia@ifpan.edu.pl}
}


\date{Date: October 4, 2017; submitted to Journal of Superconductivity and Novel Magnetism, DOI: \href{http://dx.doi.org/10.1007/s10948-017-4366-0}{10.1007/s10948-017-4366-0}}

\maketitle

\begin{abstract}
The paper is devoted to a study of superconducting properties of population-imbalanced fermionic mixtures in {\it quasi}-one dimensional optical lattices.
The system can be effectively described by the attractive Hubbard model with the Zeeman magnetic field term.
We investigated the ground state phase diagram of the model as a function of the chemical potential and the magnetic field.
The ground state of the system exhibits the conventional BCS-type superconductivity as well as the unconventional so-called Fulde-Ferrell-Larkin-Ovchinnikov state, in which the total momentum of Cooper pairs is non-zero.
We determine the orders of transitions as well as the behavior of order parameters with a change of the model parameters.
\end{abstract}

\section{Introduction}
\label{sec.intro}

The unconventional superconductivity is still very fascinating but unsolved problem.
It involves such issues as superconductivity with extremely short coherence length \cite{Robaszkiewicz.81,Robaszkiewicz.82,Dutta.15,Micnas.91}, the BCS-BEC crossover \cite{Micnas.91,Bourdel.04,cichy.14}, unconventional pairings (particularly those with non-zero total momentum of Cooper pairs, ${\bm Q}\neq0 $, e.g.  the so-called Fulde-Ferrell-Larkin-Ovchinnikov state -- FFLO, \cite{Fulde.64,Larkin.65,jakubczyk.17}).
The phenomenon of the FFLO superconductivity can be realized possibly in many physical systems:
(i) in condensed matter physics, e.g. iron-based superconductors \cite{Zocco.13,ptok.crivelli.13,fese.14,cho.yang.17,Ptok.Kapcia.17}, heavy-fermion compounds \cite{Ptok.Kapcia.17,Bianchi.02,Bianchi.03,Matsuda.07,Kenzelmann.08,maska.mierzejewski.10,kaczmarczyk.spalek.10}, and organic conductors \cite{Uji.06,Lortz.07,Mayaffre.14}, as well as
(ii) in ultra cold atomic gases on optical lattices \cite{Liu.03,Hu.06,Guan.07,Orso.07,Hu.07,Luscher.08,Wolak.12,Guan.13,Pahl.04,cichy.cichy.15,Hu.15}.
In the latter group of systems the interaction parameters can be changed in a wider range \cite{Fedichev.96,Schunck.07,Partridge.06,Zwierlein.06,Georgescu.14}.
Thus, it makes that group better candidate to experimentally investigate the model Hamiltonians.

Motivated by the experimental feasibility of such systems with ultracold gases loaded on a {\it quasi}-one dimensional lattice, we studied the unconventional superfluid phases of the attractive Hubbard model, in the presence of an external magnetic field \cite{ptok.cichy.17}.
In that paper, we have shown that the system evolves from the BCS-type superconducting state (at small field) to the FFLO phase (for sufficiently large field).
In an extremal case, the momentum of Cooper pairs can lie on the vertex of the first Brillouin zone and the so-called $\eta$ phase emerges.
In this work we present and discuss in details the dependence (as a function of chemical potential and magnetic field) of the following quantities:
(i) an amplitude of superconducting order parameter,
(ii) a total momentum of Cooper pairs, and
(iii) the particle concentration.

The rest of the paper is organised as follows.
In Section~\ref{sec.model} the investigated model is presented and the method of its solution is briefly discussed.
Section~\ref{sec.num} is devoted to a discussion of the ground state phase diagram of the model, particularly focussing on changes of order parameters with the model parameters.
Finally, in Section~\ref{sec.sum} we summarize the results of the present work.

\section{Model and methods}
\label{sec.model}

In this paper, we study a one-dimensional chain with a BCS-type superconducting pairing term (i.e. {\it s-wave} one).
The system is described by the attractive Hubbard model ($U<0$) in a magnetic field \cite{ptok.cichy.17}, which in the real space can be written in the following form:
\begin{eqnarray}
\label{ham_real}
\hat{\mathcal{H}} = \sum_{ \langle i,j \rangle \sigma } \left( - t - (\mu + \sigma h ) \delta_{ij} \right) \hat c_{i\sigma}^{\dagger} \hat c_{j\sigma} + U \sum_{i} \hat n_{i\uparrow} \hat n_{i\downarrow},
\end{eqnarray}
where $\hat{c}^\dag_{\sigma}$ ($\hat{c}_{\sigma}$) denotes an operator of creation (annihilation) of the electron with spin $\sigma \in \{ \uparrow , \downarrow \}$ at site $i$ and $\hat{n}_{i\sigma} = \hat{c}^\dagger_{i\sigma} \hat{c}_{i\sigma} $ is particle number operator.
$t>0$ is the hopping between the nearest-neighbor sites,  $U < 0$ is the on-site pairing interaction.
$\mu$ is the chemical potential, which determines the average number of particles $n = (1 / N) \sum_{i\sigma} \langle \hat{n}_{i\sigma} \rangle$ in the system (filling) ($N$ is the total number of sites in the lattice).
Finally, $h$ is a Zeeman field, which can originate from an external magnetic field (in $g \mu_B \slash 2$ units) or from population imbalance in the context of the cold atomic Fermi gases.
Moreover, we can introduce $\mu_{\sigma} = \mu + \sigma h$ as the effective chemical potential of atoms with (pseudo) spin $\sigma$.
The second term of Hamiltonian (\ref{ham_real}) is decoupled within the mean-field approximation, which takes into account only superconducting averages:
\begin{eqnarray}
\hat n_{i\uparrow} \hat n_{i\downarrow} = \Delta_{i}^{\ast} \hat c_{i\downarrow} \hat c_{i\uparrow} + \Delta_{i} \hat c_{i\uparrow}^{\dagger} \hat c_{i\downarrow}^{\dagger} - | \Delta_{i} |^{2},
\end{eqnarray}
where $\Delta_{i} = \langle \hat c_{i\downarrow} \hat c_{i\uparrow} \rangle$ is  the site-dependent on-site superconducting order parameter (SOP).
Then, the mean-field Hamiltonian in the real space is written in the form:
\begin{eqnarray}
\label{ham_MF}
\hat{\mathcal{H}}^{MF} &=& \sum_{ \langle i,j \rangle \sigma } \left( - t - ( \mu + \sigma h ) \delta_{ij} \right) \hat c_{i\sigma}^{\dagger} \hat c_{j\sigma} \\
\nonumber &+& U \sum_{i} \left( \Delta_{i}^{\ast} \hat c_{i\downarrow} \hat c_{i\uparrow} + H.c. \right) - U \sum_{i} | \Delta_{i} |^{2}.
\end{eqnarray}
Without loss of generality, one can write down the SOP as: $\Delta_{i} = \Delta_{0} \exp ( i {\bm Q} \cdot {\bm R}_{i} )$, where $\Delta_0>0$ is the spatially oscillating amplitude of the SOP and ${\bm Q}$ is the total momentum of the Cooper pairs.
We assume that the lattice constant is equal to one, i.e. $a=1$.
In a one-dimensional case considered here, $|{\bm Q}| = Q_x$, where $Q_x$ is the (absolute value of) coordinate (the only one) of the vector ${\bm Q}$ (with the largest allowed value $Q_x^{max}\equiv\pi$).
The procedure of numerical solving of the system and final equations for the grand canonical potential and order parameters are presented in Ref. \cite{ptok.cichy.17,januszewski.ptok.15}.
Notice that all found solutions correspond to minimal value of the grand canonical potential (with respect to $\Delta_0$ and ${\bm Q}$) at fixed model parameters \cite{ptok.cichy.17,januszewski.ptok.15}.
Below, we just discuss the behavior of the quantities in the system for some exemplary value of the on-site attraction.

\section{Numerical results and discussion}
\label{sec.num}

\begin{figure}
	\begin{center}
		\includegraphics[width=0.45\textwidth]{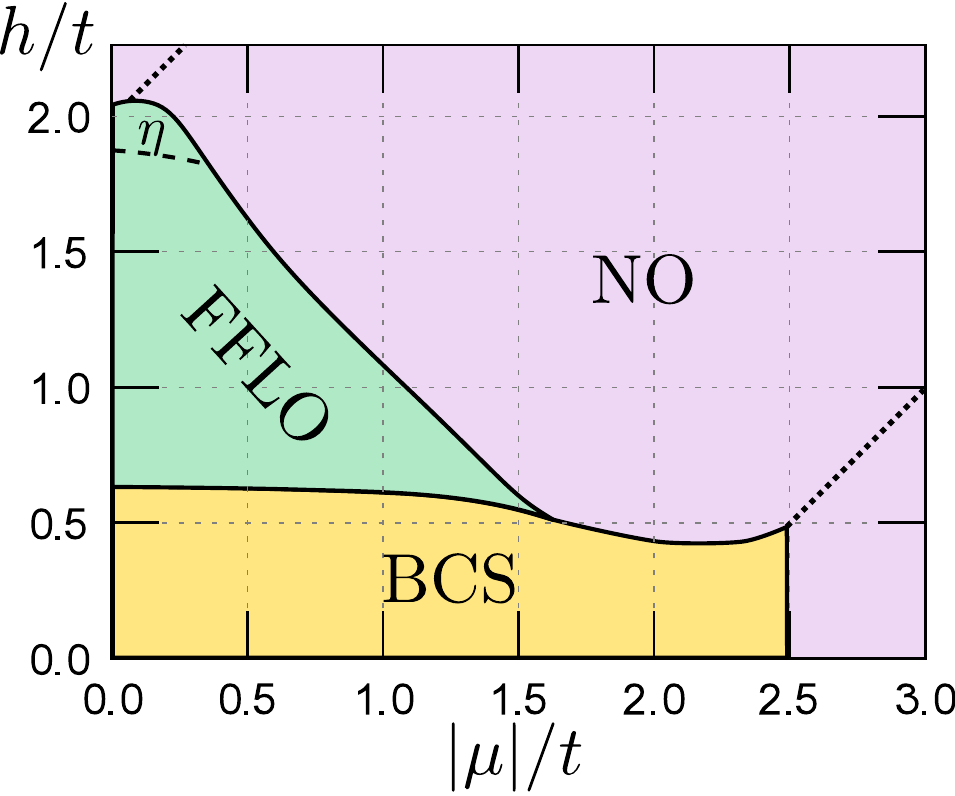}
	\end{center}
	\caption{
		The phase diagram of the model for $U/t=-3.0$ as a function of $|\mu|/t$ and $h/t$.
		BCS denotes the usual superconducting BCS-type {\it s-wave} state, FFLO labels the polarized superconducting state with non-zero momentum of Cooper pairs, whereas NO corresponds to normal (non-ordered) phase.
		Additionally, inside the FFLO region above the dashed line, there is a region, where the $\eta$ phase is distinguished.
		Inside the NO area, the NO phase with $n=1$ (for small $|\mu|/t$ and large $h/t$) and the NO phase with $|n-1|=1$ (for large $|\mu|/t$ and small $h/t$) are indicated by the dotted lines.
	}
	\label{fig:df}
\end{figure}

In this section, we consider the system with  a value of the pairing interaction set as $U/t=-3$.
The ground state phase diagram of model (\ref{ham_MF}) is presented in Fig.~\ref{fig:df}.
In the absence of an external Zeeman field, the usual superconducting BCS-type {\it s-wave} state (with $\Delta_0\neq0$ and $Q_x=0$) is stable.
As the magnetic field increases, superfluidity is destroyed due to paramagnetic effects or by population imbalance.
Hence, the unpolarized BCS-like superconducting phase undergoes a first order phase transition to the polarized normal  (NO) state (with $\Delta_0=0$, $Q_x$ -- undetermined, , and $|n-1| \neq 1$) or to the FFLO phase (with $\Delta_0\neq0$ and $Q_x\neq0$).
Inside the FFLO phase, for large fields and near half-filling, the  $\eta$-FFLO superconducting phase (i.e. the FFLO phase with $Q_x=Q_x^{max}$) is also identified.
Increasing higher the field and close to half-filling, the FFLO-$\eta$-pairing superconducting phase undergoes a first-order phase transition to the normal state  (this transition can be also second-order for larger $|\mu|/t$, see the text below).
Moreover, for magnetic fields larger than the top of the band (i.e. $2t$ for one-dimensional case studied here), one can find the magnetic Lifshitz transition~\cite{ptok.cichy.17,ptok.kapcia.17}.
As a consequence of the relatively strong pairing interaction, the superconductivity still exists in the system, even if the Fermi surface for one type spin disappears.

Now we discuss the evolution of
(i) the amplitude $\Delta_0$ of the superconducting order parameter,
(ii) the component $Q_x$ of the total momentum of Cooper pairs, and
(iii) electron concentration $n$ in the system as a function of the model parameters $\mu$ and $h$.

\begin{figure}
	\begin{center}
		\includegraphics[width=0.48\textwidth]{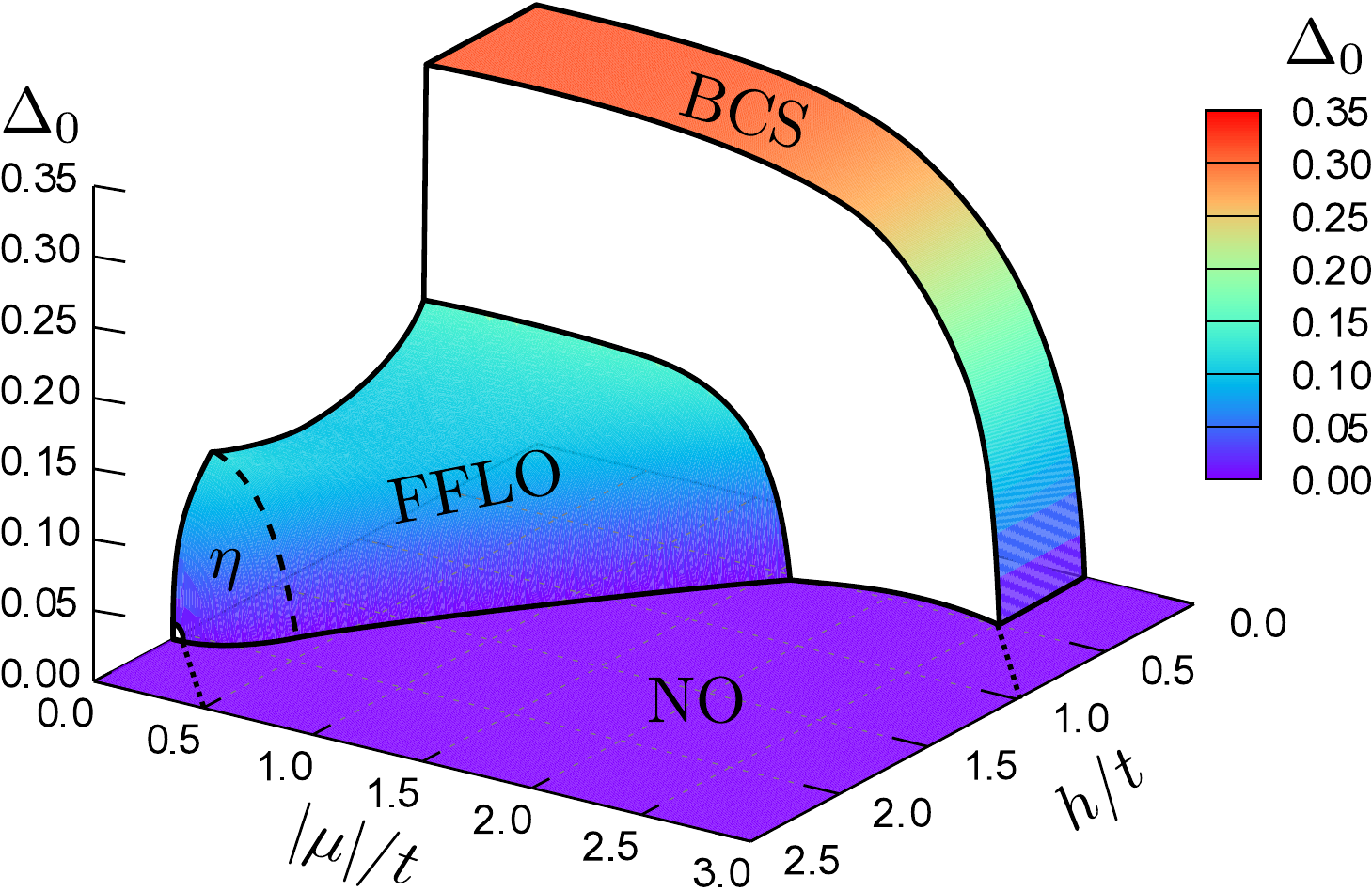}
	\end{center}
	\caption{%
		The amplitude $\Delta_0$ of the superconducting order parameter for $U/t=-3.0$ as a function of $|\mu|/t$ and $h/t$.
		The white planes denote discontinuous changes of $\Delta_0$ at the boundaries between phases (notice the small region at the $\eta$-FFLO--NO boundary  near $|\mu|/t=0$).
	}
	\label{fig:gap}
\end{figure}

Fig.~\ref{fig:gap} presents the amplitude $\Delta_0$ of the superconducting order parameter.
In the BCS phase $\Delta_0$ is monotonously decreasing function of $|\mu|/t$ and it is independent of $h/t$.
At the transition between the BCS phase and the NO phase with $n=0$ the $\Delta_0$ vanishes continuously as it should behave at continuous transition.
At $h/t\neq0$ the BCS--NO transition is discontinuous (with discontinuous change of $\Delta_0$).
Similarly, at the BCS--FFLO boundary $\Delta_0$ exhibits a discontinuous change to a lower value.
Inside the FFLO region (including also $\eta$ phase) $\Delta_0$ decreases monotonously with increasing $|\mu|/t$ and $h/t$.
At the FFLO--NO boundary parameter $\Delta_0$ vanishes continuously to $\Delta_0=0$
The transition between the $\eta$-FFLO phase and the NO phase is continuous only in some range of model parameters.
For high magnetic field and near the half-filling it is associated to a discontinuity of $\Delta_0$ (it is hardly visible in Fig.~\ref{fig:gap}).

\begin{figure}
	\begin{center}
		\includegraphics[width=0.48\textwidth]{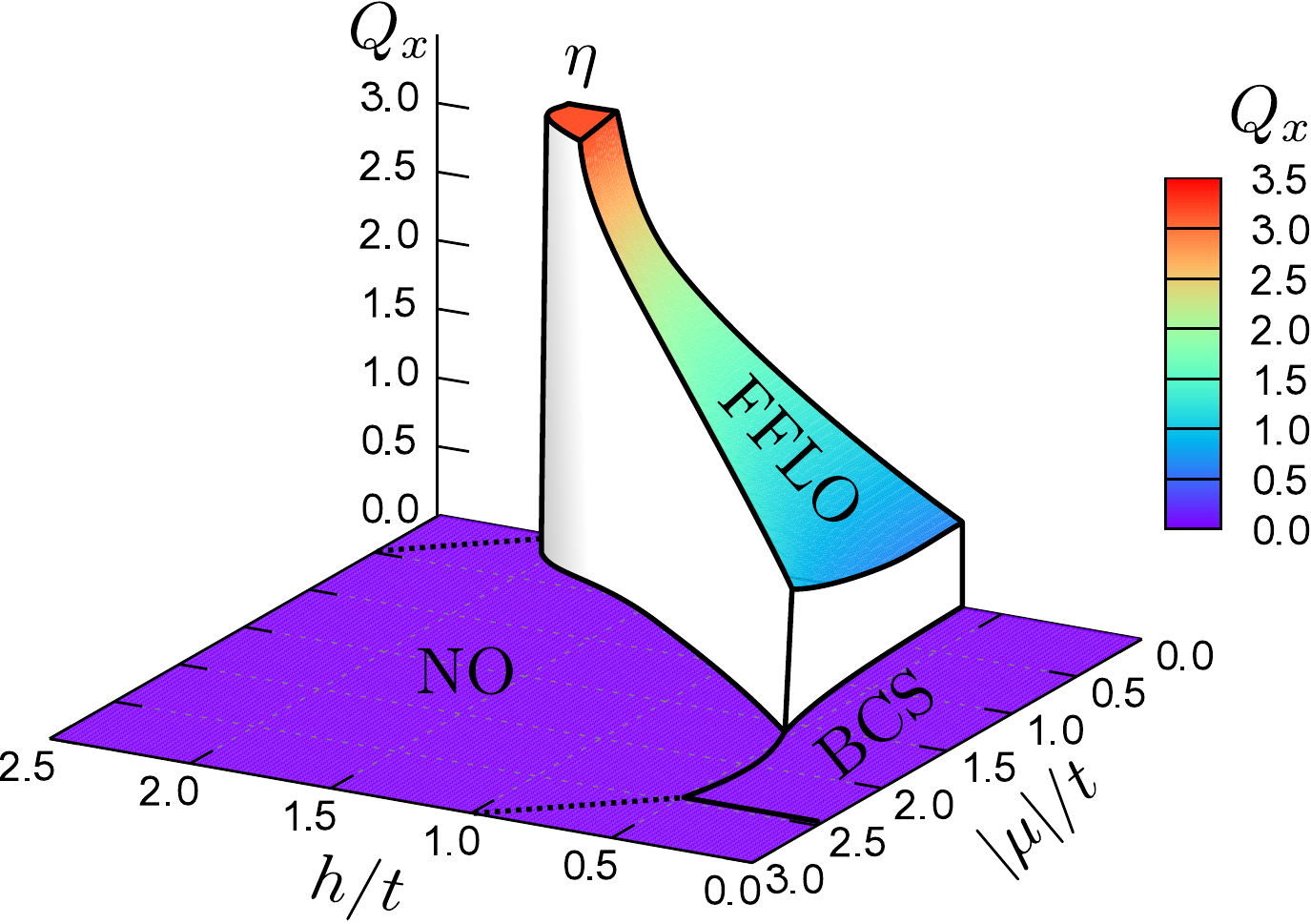}
	\end{center}
	\caption{%
		The momentum  $Q_x$ of superconducting pairs for $U/t=-3.0$ as a function of $|\mu|/t$ and $h/t$.
		At the $\eta$-FFLO phase $Q_x=\pi$.
		In the NO region $Q_x$ is undetermined (not well defined), but in the figure we have adopted the convention that in the NO phase $Q_x=0$.
        The white planes denote discontinuous changes of $Q_x$ at the boundaries between phases.
	}
	\label{fig:qx}
\end{figure}

The ground state values of $Q_x$ are presented in Fig.~\ref{fig:qx}.
As one can expect, $Q_x=0$ at the BCS phase, where the total momentum of Copper pairs is zero.
In the FFLO phase $Q_x\neq0$ and it increases with increasing of the magnetic field to its maximal value $Q_x^{max}=\pi$.
In the whole region of $\eta$-FFLO phase, the total momentum of the pairs does not change and equals $Q_x=Q_x^{max}$.
Notice that at the BCS--FFLO boundary $Q_x$ changes discontinuously.
$Q_x$ is continuous at the boundary between the $\eta$-FFLO phase (with $Q_x=Q_x^{max}$) and the FFLO phase (where $0<Q_x<Q_x^{max}$).
At the NO region $Q_x$ is undetermined (i.e. it is not well-defined, no Cooper pairs in the system), but in Fig.~\ref{fig:qx} we have adopted the convention that in the NO phase $Q_x=0$.

\begin{figure}
	\begin{center}
		\includegraphics[width=0.48\textwidth]{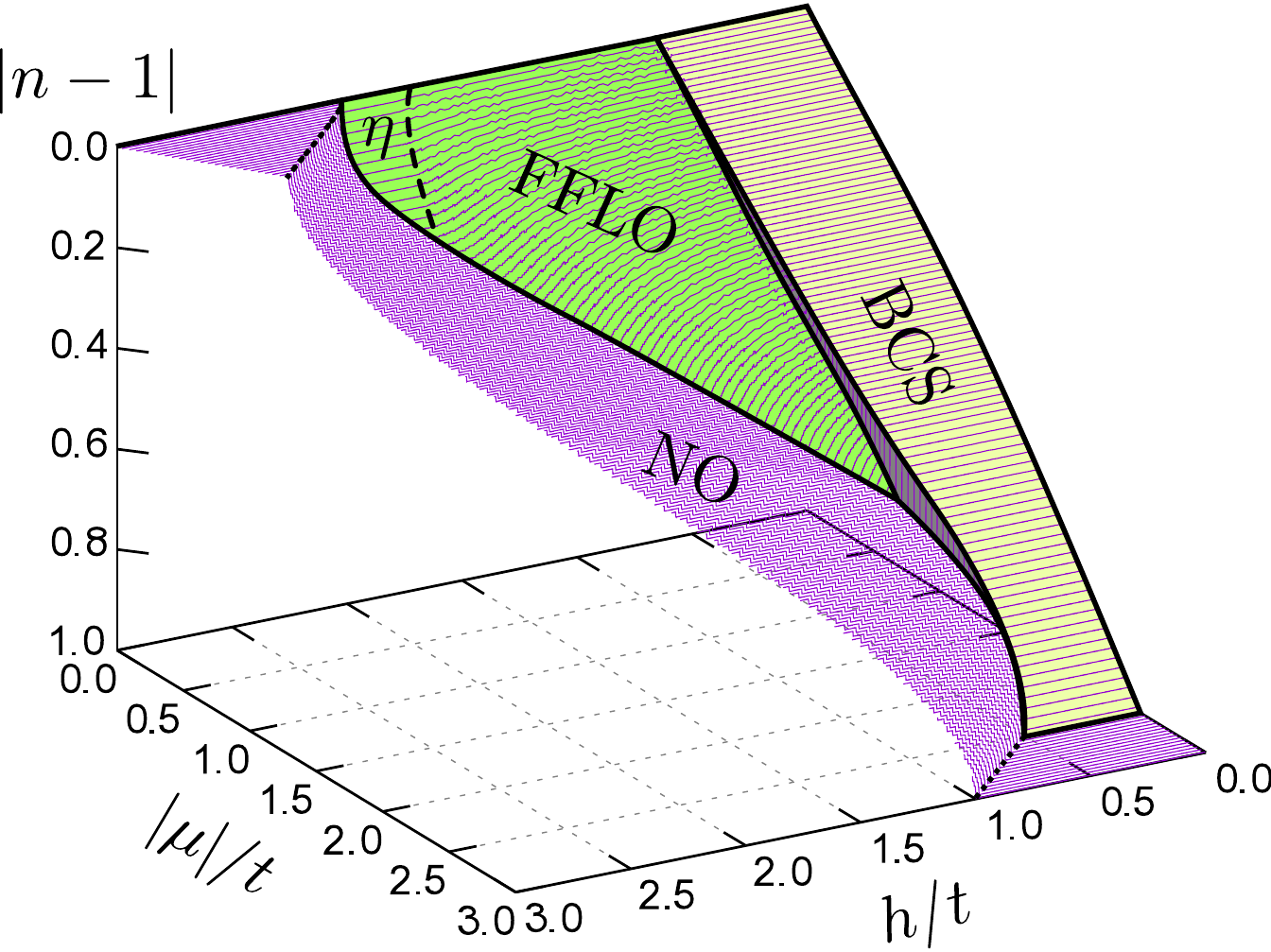}
	\end{center}
	\caption{%
		The average concentration $|n-1|$ of the particles for fixed $U/t=-3.0$ as a function of $|\mu|/t$ and $h/t$.
		The discontinuities of $|n-1|$ occur at the same phase boundaries as those for $\Delta_0$, but they are less visible in the figure. 
		Notice that $z$-axis of the plot (correponding to $|n-1|$ quantity) has inverted direction.
	}
	\label{fig:nn}
\end{figure}

The very important feature is the behavior of the particle concentration $n$, particularly in the context of the phase separations (cf. Refs. \cite{Arrigoni.Strinati.91,bak.04,kapcia.robaszkiewicz.12,kapcia.czart.16} and references therein).
If $n$ changes discontinuously from $n_-$ to $n_+$ at the boundary line between two phases in the diagram for fixed $\mu$, the phases can co-exist on the phase diagram as a function of $n$.
The dependence of particle concentration $n$ (precisely the value of $|n-1|$) with changing the model parameters is shown in Fig.~\ref{fig:nn}.
At fixed $h/t$, the value of $|n-1|$ is increasing function of $|\mu|/t$.
It turns out that the discontinuous changes of $n$ occur at the same phase boundaries where $\Delta_0$ exhibits discontinuity.
Namely, the BCS--FFLO, BCS--NO (only to NO with $n\neq0$) and $\eta$-FFLO--NO (only for high magnetic field and near the half-filling) transitions are associated to abrupt changes of $n$.
The value of $|n-1|$ in the FFLO phase is smaller than those in the NO phase (at the FFLO--NO boundary).
Similarly,  at the BCS--FFLO boundary the value of $|n-1|$ changes discontinuously to higher value in the FFLO phase.
At the discontinuous BCS--NO boundary the relative values of concentrations depend on the value of $\mu/t$.
For the discontinuous $\eta$-FFLO--NO transition the lower value of $|n-1|$ is in the NO phase.
As a result one can distinguish three different phase separated states, which can occur in the phase diagram as a function of $n$:
phase separation between the BCS and FFLO phases,
phase separation between the BCS and NO phases, and finally phase separation between the $\eta$-FFLO and NO phases (cf. also \cite{ptok.cichy.17}).

The behavior of  $\Delta_0$, $Q_x$ and $n$ should not strongly depend on low and intermediate values of $U<0$ qualitatively since it only determines the magnitude of the pairing potential in the system.
Notice also that the ground state phase diagrams for different attractive $U$ presented in Ref.~\cite{ptok.cichy.17} do not modify qualitatively with changing $U$ in this range of $U$.

\section{Summary}
\label{sec.sum}

In this work we have studied the ground state of the attractive Hubbard model focusing on the behavior of order parameters.
We found that in the ground state of the system the following phases can occur: the NO phase, the BCS phase and the FFLO phase (with its extreme case --- the $\eta$-FFLO phase).
The phase transition between the BCS and FFLO-phase is discontinuous one, whereas the FFLO--NO and FFLO--$\eta$-FFLO transitions are continuous.
The BCS--NO and $\eta$-FFLO--NO transitions can be of both types, depending on the region of the phase diagram.

Notice that the mean-field approximation is generally valid only for small $U$ and high-dimensions.
It overestimates usually critical temperatures and the range of stability of the phases with long-range order.
However, the mean-field approach gives at least qualitative description of the system in the ground state, even in the strong coupling limit \cite{Micnas.91}.
Nevertheless, we used this approach to study {\it quasi}-one dimensional model describing fermionic ultracold gas on the optical lattice.
According to the Mermin-Wagner theorem, a one-dimensional superfluid system cannot support superfluidity and would possess, at best, algebraically decaying long-range order at zero temperature \cite{Mermin.66} (cf. also with Refs.~\cite{Lieb.89,Uhrig.89,Muller.89}).
The real systems of atoms on the optical lattices are {\it quasi}-one dimensional systems which means that the one-dimensional cylindrical-shaped regions are weakly coupled with each other, what makes the mean-field approximation more appropriate.

\begin{acknowledgements}
The authors thank Jan Bara\'nski,  Krzysztof Cichy, Anna Ciechan, Tadeusz Doma\'{n}ski, and Matteo Rizzi for valuable comments and discussions.
This work was partially supported by the National Science Centre (NCN, Poland) under grants no.
UMO-2016/20/S/ST3/00274 (A.P.),
UMO-2017/24/C/ST3/00357 (A.C.),
UMO-2016/21/D/ST3/03385 (K.J.K.),
and
UMO-2017/24/C/ST3/00276 (K.J.K.).
K.R. acknowledges the support from CIBioFi and the Colombian Science, Technology and Innovation Foundation--COLCIENCIAS ``Francisco Jos\'{e} de Caldas'' under project 1106-712-49884 (contract No.264-2016) and --General Royalties System (Fondo CTeI-SGR) under contract No. BPIN 2013000100007.
\end{acknowledgements}

\small{

}


\begin{thebibliography}{10}

\bibitem{Robaszkiewicz.81}
Robaszkiewicz, S., Micnas, R., Chao, K.A.:
\arttitle{Hartree theory for the negative-$U$ extended Hubbard model: Ground state}
Phys. Rev. B \textbf{24}, 4018 (1981)
\linkDOIOK{10.1103/PhysRevB.24.4018}

\bibitem{Robaszkiewicz.82}
Robaszkiewicz, S., Micnas, R., Chao, K.A.:
\arttitle{Hartree theory for the negative-$U$, extended Hubbard model. II. Finite temperature}
Phys. Rev. B \textbf{26}, 3915 (1982)
\linkDOIOK{10.1103/PhysRevB.26.3915}

\bibitem{Dutta.15}
Dutta, O., Gajda, M., Hauke, P., Lewenstein, M.,  L\"uhmann, D.-S.,  Malomed, B.A., Sowi\'nski, T.,  Zakrzewski, J.:
\arttitle{Non-standard Hubbard models in optical lattices: A review}
Rep. Prog. Phys. \textbf{78}, 066001 (2015)
\linkDOIOK{10.1088/0034-4885/78/6/066001}

\bibitem{Micnas.91}
Micnas, R., Ranninger, J., Robaszkiewicz, S.:
\arttitle{Superconductivity in narrow-band systems with local nonretarded attractive interactions}
Rev. Mod. Phys. \textbf{62}, 113 (1990)
\linkDOIOK{10.1103/RevModPhys.62.113}

\bibitem{Bourdel.04}
Bourdel, T., Khaykovich, L.,  Cubizolles, J.,  Zhang, J.,  Chevy, F., Teichmann, M., Tarruell, L., Kokkelmans,  S.J.J.M.F., Salomon,  C.,
\arttitle{Experimental Study of the BEC-BCS Crossover Region in Lithium 6}
Phys. Rev. Lett. \textbf{93}, 050401 (2004)
\linkDOIOK{10.1103/PhysRevLett.93.050401}

\bibitem{cichy.14}
Cichy, A., Micnas, R.:
\arttitle{The spin-imbalanced attractive Hubbard model in: Phase diagrams and BCS-BEC crossover at low filling}
Ann. Phys. \textbf{347}, 207 (2014).
\linkDOIOK{10.1016/j.aop.2014.04.014}

\bibitem{Fulde.64}
Fulde, P., Ferrell, R.A.:
\arttitle{Superconductivity in a strong spinexchange field}
Phys. Rev. \textbf{135}, A550 (1964)
\linkDOIOK{10.1103/PhysRev.135.A550}

\bibitem{Larkin.65}
Larkin, A.I.,  Ovchinnikov, Yu.N.:
\arttitle{Nonuniform state of superconductors}
Zh. Eksp. Teor. Fiz. \textbf{47}, 1136 (1964) [Sov. Phys. JETP \textbf{20}, 762 (1965)].

\bibitem{jakubczyk.17}
Jakubczyk, P.:
\arttitle{Renormalization theory for the Fulde-Ferrell-Larkin-Ovchinnikov states at $T > 0$}
Phys. Rev. A \textbf{95}, 063626 (2017)
\linkDOIOK{10.1103/PhysRevA.95.063626}

\bibitem{Zocco.13}
Zocco, D.A., Grube, K., Eilers, F., Wolf, T.,  von L\"ohneysen, H.:
\arttitle{Pauli-Limited Multiband Superconductivity in KFe$_2$As$_2$}
Phys. Rev. Lett. \textbf{111}, 057007 (2013)
\linkDOIOK{10.1103/PhysRevLett.111.057007}

\bibitem{ptok.crivelli.13}
Ptok, A.,  Crivelli, D.:
\arttitle{The Fulde-Ferrell-Larkin-Ovchinnikov State in Pnictides}
J. Low Tem. Phys. \textbf{172}, 226 (2013)
\linkDOIOK{10.1007/s10909-013-0871-0}

\bibitem{fese.14}
Kasahara, S., Watashige, T., Hanaguri, T., Kohsaka, Y., Yamashita, T., Shimoyama, Y., Mizukami, Y., Endo, R., Ikeda, H., Aoyama, K., Terashima, T., Taichi, S., Wolf, T., von L\"{o}hneysen, H., Shibauchi, T., Matsuda, Y.:
\arttitle{Field-induced superconducting phase of FeSe in the BCS-BEC cross-over}
PNAS \textbf{111}, 16309 (2014)
\linkDOIOK{10.1073/pnas.1413477111}

\bibitem{cho.yang.17}
Cho, Ch.-w., Yang, J. H., Shen, J., Wolf, T., Lortz,  R.:
\arttitle{Thermodynamic evidence for the Fulde-Ferrell-Larkin-Ovchinnikov state in the KFe$_2$As$_2$ superconductor}
arXiv:1708.05526

\bibitem{Ptok.Kapcia.17}
Ptok, A., Kapcia, K.J., Piekarz, P., Ole\'s, A.M.:
\arttitle{The ab initio study of unconventional superconductivity in CeCoIn$_5$ and FeSe}
New. J. Phys. \textbf{19}, 063039 (2017)
\linkDOIOK{10.1088/1367-2630/aa6d9d}

\bibitem{Bianchi.02}
Bianchi, A., Movshovich, R., Oeschler, N., Gegenwart, P., Steglich, F., Thompson, J.D., Pagliuso, P.G., Sarrao,  J.L.:
\arttitle{First-Order Superconducting Phase Transition in CeCoIn$_5$}
Phys. Rev. Lett. \textbf{89}, 137002 (2002)
\linkDOIOK{10.1103/PhysRevLett.89.137002}

\bibitem{Bianchi.03}
Bianchi, A., Movshovich, R., Capan,  C., Pagliuso,  P.G., Sarrao, J.L.:
\arttitle{Possible Fulde-Ferrell-Larkin-Ovchinnikov Superconducting State in CeCoIn$_5$}
Phys. Rev. Lett. \textbf{91}, 187004 (2003)
\linkDOIOK{10.1103/PhysRevLett.91.187004}

\bibitem{Matsuda.07}
Matsuda, Y., Shimahara,  H.:
\arttitle{Fulde-Ferrell-Larkin-Ovchinnikov state in heavy fermion superconductors},
J. Phys. Soc. Jpn. \textbf{76}, 051005 (2007)
\linkDOIOK{10.1143/JPSJ.76.051005}

\bibitem{Kenzelmann.08}
Kenzelmann, M., Str\"assle, Th., Niedermayer, Ch., Sigrist, M., Padmanabhan, B., Zolliker, M., Bianchi,  A.D., Movshovich, R., Bauer, E.D., Sarrao, J. L., Thompson,  J.D.:
\arttitle{Coupled superconducting and magnetic order in CeCoIn$_5$}
Science \textbf{321}, 1652 (2008)
\linkDOIOK{10.1126/science.1161818}

\bibitem{maska.mierzejewski.10}
Ma\'ska, M.M., Mierzejewski, M., Kaczmarczyk,  J.,  Spa\l{}ek, J.:
\arttitle{Superconducting Bardeen-Cooper-Schrieffer versus Fulde-Ferrell-Larkin-Ovchinnikov states of heavy quasiparticles with spin-dependent masses and Kondo-type pairing}
Phys. Rev. B \textbf{82}, 054509 (2010)
\linkDOIOK{10.1103/PhysRevB.82.054509}

\bibitem{kaczmarczyk.spalek.10}
Kaczmarczyk,  J.,  Spa\l{}ek, J.:
\arttitle{Unconventional superconducting phases in a correlated two-dimensional Fermi gas of nonstandard quasiparticles: a simple model}
J. Phys.: Condens. Matter \textbf{22}, 355702 (2010)
\linkDOIOK{10.1088/0953-8984/22/35/355702}

\bibitem{Uji.06}
Uji, S., Terashima, T., Nishimura, M., Takahide, Y., Konoike, T., Enomoto, K., Cui, H., Kobayashi, H., Kobayashi, A., Tanaka, H., Tokumoto, M., Choi, E.S., Tokumoto, T., Graf, D., Brooks,  J.S.:
\arttitle{Vortex Dynamics and the Fulde-Ferrell-Larkin-Ovchinnikov State in aMagnetic-Field-Induced Organic Superconductor}
Phys. Rev. Lett. \textbf{97}, 157001 (2006)
\linkDOIOK{10.1103/PhysRevLett.97.157001}

\bibitem{Lortz.07}
Lortz, R., Wang, Y., Demuer, A., B\"ottger, P.H.M., Bergk, B., Zwicknagl, G., Nakazawa, Y., Wosnitza,  J.:
\arttitle{Calorimetric Evidence for a Fulde-Ferrell-Larkin-Ovchinnikov Superconducting State in the Layered Organic Superconductor $\kappa$-(BECT-TTF)$_2$Cu(NCS)$_2$}
Phys. Rev. Lett. \textbf{99}, 187002 (2007)
\linkDOIOK{10.1103/PhysRevLett.99.187002}

\bibitem{Mayaffre.14}
Mayaffre, H., Kramer,  S., Horvatic, M., Berthier, C., Miyagawa, K., Kanoda, K., Mitrovic,  V.F.:
\arttitle{Evidence ofAndreev bound states as a hallmark of the FFLO phase in $\kappa$-(BEDT-TTF)$_2$Cu(NCS)$_2$}
Nat. Phys. \textbf{10}, 928 (2014)
\linkDOIOK{10.1038/nphys3121}

\bibitem{Liu.03}
Liu, W.V., Wilczek, F.:
\arttitle{Interior Gap Superfluidity}
Phys. Rev. Lett. \textbf{90}, 047002 (2003)
\linkDOIOK{10.1103/PhysRevLett.90.047002}

\bibitem{Hu.06}
Hu, H.,  Liu, X.-J.:
\arttitle{Mean-field phase diagrams of imbalanced Fermi gases near a Feshbach resonance}
Phys. Rev. A \textbf{73}, 051603 (2006)
\linkDOIOK{10.1103/PhysRevA.73.051603}

\bibitem{Guan.07}
Guan, X. W., Batchelor, M.T., Lee, C., Bortz, M.:
\arttitle{Phase transitions and pairing signature in strongly attractive Fermi atomic gases}
Phys. Rev. B \textbf{76}, 085120 (2007)
\linkDOIOK{10.1103/PhysRevB.76.085120}

\bibitem{Orso.07}
Orso, G.:
\arttitle{Attractive Fermi Gases with Unequal Spin Populations in Highly Elongated Traps}
Phys. Rev. Lett. \textbf{98}, 070402 (2007)
\linkDOIOK{10.1103/PhysRevLett.98.070402}

\bibitem{Hu.07}
Hu, H., Liu, X.-J., Drummond, P.D.:
\arttitle{Phase Diagram of a Strongly Interacting Polarized Fermi Gas in One Dimension}
Phys. Rev. Lett. \textbf{98}, 070403 (2007)
\linkDOIOK{10.1103/PhysRevLett.98.070403}

\bibitem{Luscher.08}
L\"uscher, A., Noack, R.M., L\"auchli,  A.M.:
\arttitle{Fulde-Ferrell-Larkin-Ovchinnikov state in the one-dimensional attractive Hubbard model and its fingerprint in spatial noise correlations}
Phys. Rev. A \textbf{78}, 013637 (2008)
\linkDOIOK{10.1103/PhysRevA.78.013637}

\bibitem{Pahl.04}
Pahl S., Koinov, Z.:
\arttitle{Phase diagram of a $^6$Li-$^{40}$K mixture in a square lattice}
J. Low Temp. Phys. \textbf{176}, 113 (2014)
\linkDOIOK{10.1007/s10909-014-1166-9}

\bibitem{Wolak.12}
Wolak, M. J., Gr\'emaud, B., Scalettar, R.T., Batrouni,  G.G.:
\arttitle{Pairing in a two-dimensional Fermi gas with population imbalance}
Phys. Rev. A \textbf{86}, 023630 (2012)
\linkDOIOK{10.1103/PhysRevA.86.023630}

\bibitem{Guan.13}
Guan, X.-W., Batchelor,  M.T., Lee, Ch.:
\arttitle{Fermi gases in one dimension: From Bethe ansatz to experiments}
Rev. Mod. Phys. \textbf{85}, 1633 (2013)
\linkDOIOK{10.1103/RevModPhys.85.1633}

\bibitem{cichy.cichy.15}
Cichy, A., Cichy,  K., Polak, T.P.:
\arttitle{Competition between Abelian and Zeeman magnetic field effects in a two-dimensional ultracold gas of fermions}
Ann. Phys. \textbf{354}, 89 (2015)
\linkDOIOK{10.1016/j.aop.2014.12.008}

\bibitem{Hu.15}
Hu, A., Ma\'ska,  M.M., Clark, Ch.W., Freericks, J.K.:
\arttitle{Robust finite-temperature disordered Mott-insulating phases in inhomogeneous Fermi-Fermi mixtures with density and mass imbalance}
Phys. Rev. A \textbf{91}, 063624 (2015)
\linkDOIOK{10.1103/PhysRevA.91.063624}

\bibitem{Fedichev.96}
Fedichev, P.O., Kagan, Yu., Shlyapnikov, G.V., Walraven,  J.T.M.:
\arttitle{Influence of Nearly Resonant Light on the Scattering Length in Low-Temperature Atomic Gases}
Phys. Rev. Lett. \textbf{77}, 2913 (1996)
\linkDOIOK{10.1103/PhysRevLett.77.2913}

\bibitem{Partridge.06}
Partridge, G.B., Li, W., Kamar,  R. I., Liao, Y.-a.,  Hulet, R.G.:
\arttitle{Pairing and phase separation in a polarized Fermi gas}
Science \textbf{311}, 503 (2006)
\linkDOIOK{ 10.1126/science.1122876}

\bibitem{Zwierlein.06}
Zwierlein, M.W., Ketterle, W.:
\arttitle{Comment on ``Pairing and phase separation in a polarized Fermi gas''}
Science \textbf{314}, 54 (2006)
\linkDOIOK{10.1126/science.1129812}

\bibitem{Schunck.07}
Schunck, C.H., Shin,  Y., Schirotzek, A., Zwierlein, M.W., Ketterle, W.:
\arttitle{Pairing without superfluidity: The ground state of an imbalanced Fermi mixture}
Science \textbf{316}, 867 (2007)
\linkDOIOK{10.1126/science.1140749}

\bibitem{Georgescu.14}
Georgescu, I.M., Ashhab, S., Nori, F.:
\arttitle{Quantum simulation}
Rev. Mod. Phys. \textbf{86}, 153 (2014)
\linkDOIOK{10.1103/RevModPhys.86.153}
	
\bibitem{ptok.cichy.17}
Ptok, A.,  Cichy, A., Rodr\'{i}guez, K., Kapcia, K. J.:
\arttitle{Critical behavior in one dimension: Unconventional pairing, phase separation, BEC-BCS crossover, and magnetic Lifshitz transition}
Phys. Rev. A \textbf{95}, 033613 (2017)
\linkDOIOK{10.1103/PhysRevA.95.033613}

\bibitem{januszewski.ptok.15}
Januszewski, M., Ptok, A., Crivelli, D.,  Gardas, B.:
\arttitle{GPU-based acceleration of free energy calculations in solid state physics}
Comput. Phys. Commun. \textbf{192}, 220 (2015)
\linkDOIOK{10.1016/j.cpc.2015.02.012}

\bibitem{ptok.kapcia.17}
Ptok, A., Kapcia, K.J., Cichy, A., Ole\'{s}, A.M., Piekarz, P.:
\arttitle{Magnetic Lifshitz transition and its consequences in multi-band iron-based superconductors}
Sci. Rep. \textbf{7}, 41979 (2017)
\linkDOIOK{10.1038/srep41979}

\bibitem{Arrigoni.Strinati.91}
Arrigoni, E., Strinati, G.C.:
\arttitle{Doping-induced incommensurate antiferromagnetism in a Mott-Hubbard insulator}
Phys. Rev. B \textbf{44}, 7455 (1991)
\linkDOIOK{10.1103/PhysRevB.44.7455}

\bibitem{bak.04}
B\k{a}k, M.:
\arttitle{Mixed Phase and Bound States in the Phase Diagram of the Extended Hubbard Model}
Acta Phys. Pol. A \textbf{106}, 637 (2004)
\linkDOIOK{10.12693/APhysPolA.106.637}

\bibitem{kapcia.robaszkiewicz.12}
Kapcia, K., Robaszkiewicz, S.:
\arttitle{The magnetic field induced phase separation in a model of a superconductor with local electron pairing}
J. Phys.: Condens. Matter \textbf{25}, 065603 (2013)
\linkDOIOK{10.1088/0953-8984/25/6/065603}

\bibitem{kapcia.czart.16}
Kapcia, K.J., Czart, W.R.,  Ptok, A.:
\arttitle{Phase separation of superconducting phases in the Penson-Kolb-Hubbard model}
J. Phys. Soc. Jpn. \textbf{85}, 044708 (2016)
\linkDOIOK{10.7566/JPSJ.85.044708}

\bibitem{Mermin.66}
Mermin, N.D., Wagner, H.:
\arttitle{Absence of Ferromagnetism or Antiferromagnetism in One- or Two-Dimensional Isotropic Heisenberg Models}
Phys. Rev. Lett. \textbf{17}, 1133 (1966)
\linkDOIOK{10.1103/PhysRevLett.17.1133}

\bibitem{Lieb.89}
Lieb, E.H.:
\arttitle{Two theorems on the Hubbard model}
Phys. Rev. Lett. \textbf{62}, 1201 (1989)
\linkDOIOK{10.1103/PhysRevLett.62.1201}

\bibitem{Uhrig.89}
Uhrig, G.S.:
\arttitle{Nonexistence of planar magnetic order in the one- and two-dimensional generalized Hubbard model at finite temperatures}
Phys. Rev. B \textbf{45}, 4738 (1989)
\linkDOIOK{10.1103/PhysRevB.45.4738}

\bibitem{Muller.89}
M\"uller-Hartmann, E.:
\arttitle{Correlated fermions on a lattice in high dimensions}
Z. Phys. B \textbf{74}, 507 (1989)
\linkDOIOK{10.1007/BF01311}

\end{thebibliography}
\end{document}